\documentstyle[aps,prd,preprint]{revtex}
\tighten
\begin{document}

\title{\bf Black holes are one-dimensional}

\author{Jacob D. Bekenstein\thanks{e--mail:
bekenste@vms.huji.ac.il} and Avraham E.
Mayo\thanks{e--mail:mayo@cc.huji.ac.il}}

\address{\it The Racah Institute of Physics, Hebrew University
of Jerusalem,\\ Givat Ram, Jerusalem 91904, Israel}

\maketitle

\vskip 1.0cm

\begin{abstract}
The holographic principle has revealed that physical systems in 3--D space,
black holes included, are basically  two--dimensional as far as their
information content is concerned.  This conclusion is complemented by one
sketched here: as far as entropy or information flow is concerned, a black
hole behaves as a one--dimensional channel.  We define a channel in flat
spacetime in thermodynamic terms, and contrast it with common entropy
emitting systems.  A black hole is more like the former: its entropy output
is related to the emitted power as it would be for a one-dimensional channel,
and disposal of an information stream down a black hole is limited by the
power invested in the same way as for a one-dimensional channel.
\end{abstract}

\newpage

The holographic principle\cite{thooft,susskind} claims that a
generic physical system in three spatial dimensions is fundamentally
two-dimensional.   This idea is closely connected with the
perception\cite{susskind,corley}, now commonplace, that a black hole in 3--D
space is really two-dimensional because its entropy {\it qua\/} measure of
lost information is measured by the horizon's 2--D area. In this essay we
point out a further constriction of dimensions:  viewed as an information
absorber or entropy emitter, a black hole in 3--D is fundamentally
one-dimensional, verily a portal to a one-dimensional information channel.

To show this one must define an information channel.  In flat spacetime a
channel is a complete set of unidirectionally propagating modes of
a field parametrized by a single number.  For example, all electromagnetic
modes in free space with fixed wave vector direction and particular linear
polarization constitute a channel, with the modes propagating in the
specified sense parametrized solely by frequency.  One might implement such a
channel with a straight infinitely long coaxial cable (which is well known to
transmit all frequencies) capped by polaroid filter at the entrance.  Of
course, one is not confined to electromagnetic field for sending information;
sound (acoustic field) or neutrinos will do just as well.  However, in this
essay we mostly speak in terms of photons.  A fundamental question is what is
the maximum {\it rate\/}, in quantum theory, at which information may be
transmitted in steady state down a photon channel for prescribed power $P$.
The answer was found in the 1960's\cite{lebedev}, but we reconstruct here the
much later but very simple derivation of Pendry\cite{pendry} because of its
broad applicability.

One thinks of each possible signal state as represented by a particular
occupation number state of the various propagating modes of the quantum
electromagnetic field.  Let us assume the channel is uniform in the
direction of propagation; this allows us to label the modes by momentum
$p$.  One may allow for dispersion so that a quantum with momentum $p$  has
some energy $\varepsilon(p)$. Then the propagation velocity of the quantum is
the group velocity $\upsilon(p)=d\varepsilon(p)/dp$.    According to
information theory one can identify the information rate capacity for given
$P$ with the maximal unidirectional thermodynamic entropy current that the
channel can carry for that same $P$.  This maximal entropy current obviously
occurs for the thermal state, except that we must restrict attention to modes
moving in a definite sense along the channel.  We shall leave out the factor
$\ln 2$ which translates from entropy natural units to bits.

Now the entropy $s(p) $ of any boson mode of momentum $p$ in a thermal
state (temperature $T$) is\cite{LL}
\begin{equation}
s(p)={\varepsilon(p)\over e^{\varepsilon(p)/T}-1}-\ln \left(
1-e^{-\varepsilon(p)/T}\right).
\label{s}
\end{equation}
so the entropy current in one direction is
\begin{equation}
\dot S=\int^{\infty}_0 s(p)\thinspace\upsilon(p)\thinspace dp/2\pi \hbar,
\label{current}
\end{equation}
where $dp/2 \pi \hbar $ is the number of modes per unit length in the interval
$dp$ which propagate in one direction.  This factor, when multiplied by
the group velocity, gives the unidirectional current of modes. After an
integration by parts on the second term coming from (\ref{current}), we can
cast the last result into the form
\begin{equation}
\dot S= {2\over T}\int^{\infty}_0 {\varepsilon(p)\over e^{\varepsilon(p)/T}
-1}\thinspace {d\varepsilon(p) \over dp} \thinspace {dp\over 2\pi\hbar }.
\label{entropy_flow}
\end{equation}
The first factor in the integrand is the mean energy per mode, so that the
integral represents the  unidirectional power $P$ in the channel.  Thus
\begin{equation}
\dot S=2P/T .
\label{final}
\end{equation}
The integral in Eq.~(\ref{entropy_flow}) is evaluated by cancelling the two
differentials $dp$ and assuming the energy spectrum is single valued  and
extends from 0 to $\infty$.  Then the form of the dispersion relation
$\varepsilon(p)$ does not enter and the result for the power is
\begin{equation}
P = \pi (T)^2/12 \hbar.
\label{power}
\end{equation}
Eliminating $T$ between the last two expressions gives Pendry's maximum
entropy rate for power $P$,
\begin{equation}
\dot S=(\pi P/3\hbar)^{1/2},
\label{pendry_formula}
\end{equation}

The function $\dot S(P)$ in Eq.~(\ref{pendry_formula}) is also called the
noiseless quantum channel capacity.  Surprisingly, it is independent, not
only of the form of the mode velocity $\upsilon(p)$, but also of its scale.
Thus the phonon channel capacity is as large as the photon channel capacity
despite the difference in speeds. Why? Although phonons convey information
at lower speed, the energy of a phonon is proportionately smaller than that
of a photon in the equivalent mode.  When the capacity is expressed in terms
of the energy flux, or $P$, it turns out to involve the same constants.
Formula.~(\ref{pendry_formula}) neatly characterizes what we mean by
one-dimensional transmission of entropy or information.  It refers to
transmission by use of a single species of quantum and a specific
polarization; different species and polarizations engender
separate channels.

For contrast let us derive, still in flat spacetime, the equivalent result for
the energy and entropy transmission in a single photon polarization out of a
closed hot black body surface with temperature $T$ and area $A$ into 3--D
space.  Halving the Stefan-Boltzmann law we have
\begin{equation}
P={\pi^2 T^4 A\over 120 h^3}
\label{P}
\end{equation}
as well as
\begin{equation}
\dot S =4P/3T
\label{Sdot}
\end{equation}
whereby
\begin{equation}
\dot S= {2\over 3}\left({2\pi^2AP^3\over 15\hbar^3}\right)^{1/4}
\label{3-D}
\end{equation}
Our 3--D transmission system deviates from the sleek formula
(\ref{pendry_formula}) not only in the exponent of $P$ but also in the
appearance of the measure $A$ of the system.  [In emission from a closed
curve of length $L$ in two-dimensional space the factor $ (L P^2)^{1/3}$
would appear instead of $(A P^3)^{1/4}$].  In flat spacetime we may thus
infer the dimensionality of the transmission system from the exponent of $P$
in the expression $\dot S(P)$ as well as from the value of the coefficient of
$P/T$ in expressions for $\dot S$ like (\ref{final}) or (\ref{Sdot})
[$(D+1)/D$ for
$D$ space dimensions].

Turning now to curved spacetime, the radiation from a Schwarzschild black
hole of mass $M$ in 3--D space is also given by Eqs.~(\ref{P})-(\ref{Sdot})
with $A=4\pi (2M)^2$ and $T$ the Hawking temperature $T_{\rm H}=\hbar(8\pi
M)^{-1}$, except that we must augment the expression for
$P$ by a factor $\bar\Gamma$ of order unity (Page\cite{page1} has calculated
the frequency dependent barrier transmission factor $\Gamma$ and
this must be averaged over the Planck spectrum to get $\bar\Gamma$), and
replace the $4/3$ in the expression for  $\dot S$ by another factor, $\nu$
(also calculated by Page\cite{page2}).  Eliminating
$M$ between the equations we have instead of Eq.~(\ref{3-D})
\begin{equation}
\dot S = \left({\nu^2\bar\Gamma\pi P\over
480\hbar}\right)^{1/2}.
\label{BHlimit}
\end{equation}
This looks completely different from the law (\ref{3-D}) for the hot
 closed surface in 3--D space because, unlike for the hot body, a black hole's
temperature is related to its mass in a specific way.

But (\ref{BHlimit}) {\it is\/} of the same form as Pendry's
limit (\ref{pendry_formula}) for one-channel flow.  From Page's numerical
estimates\cite{page1} we infer $\bar\Gamma=1.6267$ and take his
value\cite{page2} $\nu=1.5003$, both for a single photon polarization.  With
these the numerical coefficient of (\ref{BHlimit}) is $15.1\%$ that of
(\ref{pendry_formula}).  One consequence of the above is that it is possible
in principle to collect all the Hawking photon radiation (the
former for one polarization) by means of suitable parabolic ``mirrors'' and
``lenses'' and pipe it down a {\it single\/} straight photon channel.  This
even though the black hole emits photons in all directions and
thus, seemingly, into a gamut of channels.  Evidently in its entropy emission
properties, a black hole in 3--D space is more like a 1--D channel than like
a surface in 3--D space.  We have checked that these conclusions are not
qualitatively changed when the entropy is carried by neutrinos.

We have emphasized entropy flow out of the black hole; equally interesting is
information flow into the black hole.  One of the characteristics of a black
hole is that it acts as a sink of information.  Our results can be
construed as putting a bound on how fast information may be disposed of into a
black hole.  Let us assume we have at our disposal a certain power $P$ to
accomplish the task of getting rid of a stream of information.  We may pick
the size of black hole which suits us best.  Then by the complementary
relation between entropy and information, we may reinterpret formula
(\ref{BHlimit}) for entropy rate {\it out\/} of the black hole as also setting
a bound on the rate at which information can flow {\it into\/} the black
hole for given $P$ (there is actually a factor $\ln 2$ between them to convert
entropy natural units to bits).  The one-dimensional character of the  black
hole is central to this conclusion.  One can, of course, improve the
disposal rate by harnessing several channels (the second photon
polarization, neutrinos, etc.), but the number of these is quite limited
in nature.  One must thus pay for information disposal into a black
hole: the faster we want it done, the more power we have to put in, with
the power growing quadratically with the loss of information rate.

\noindent
{\it Acknowledgments:\hskip 0.5cm} This research was supported by
grant No. 129/00-1 of the Israel Science Foundation.

\end{document}